\documentclass[9pt,twocolumn,twoside]{opticajnl}
\journal{opticajournal} 

\setboolean{shortarticle}{true}

\usepackage{siunitx}
\usepackage{xcolor}
\usepackage[normalem]{ulem}
\usepackage[version=4]{mhchem}

\title{Sub-Doppler spectroscopy of the strontium intercombination line in a microfabricated vapor cell}

\author[1,2]{Yang Li}
\author[2]{John Kitching}
\author[2,*]{Matthew T. Hummon}

\affil[1]{Department of Physics, University of Colorado Boulder, Boulder, Colorado 80309, USA}
\affil[2]{Time and Frequency Division, National Institute of Standards and Technology, Boulder, Colorado, 80305, USA}

\affil[*]{matthew.hummon@nist.gov}

\begin{abstract}
We report a 100 kHz linewidth for the $^1$S$_0$ to $^3$P$_1$ intercombination line  of $^{88}$Sr atoms at \qty{689}{nm} in a microfabricated $\mathbf{(9 \times 14 \times 4.4 )}$~mm$^3$ vapor cell. This puts an upper bound on the residual gas pressure in the vapor cell of 10 mTorr. The microfabricatd Sr vapor cell offers an inexpensive and scalable way to lock lasers in state-of-the-art cold-atom based atomic clocks and quantum computing experiments. It also paves the way for a compact and precise optical frequency references based on alkaline earth vapor cells.
\end{abstract}

\setboolean{displaycopyright}{false} 

\begin{document}

\maketitle

\section{Introduction}
Due to their unique energy level structure, alkaline earth atoms hold a crucial position in quantum computation, information, and metrology. Laser-cooled Sr and Yb atoms are important candidates for quantum computing and information \cite{daley_quantum_2008} and are also the foundation of some of the most precise atomic clocks\cite{ludlow_optical_2015, aeppli_clock_2024}.   In these systems, the $^1$S$_0$ to $^3$P$_1$ intercombination transition provides a balance between transition strength and narrow linewidth (\qtyrange{10}{200}{\kilo\hertz}), enabling a wide range of applications including narrow-line cooling in lattice clocks \cite{ludlow_optical_2015}, and quantum state preparation and readout \cite{lis_midcircuit_2023} in alkaline earth quantum computers. Compared with cold-atom platforms, room-temperature ensembles of alkaline earth atoms offer higher potential for deployment in the field \cite{hilton_demonstration_2025} due to their far simpler implementation and more robust operation. Spectroscopy on the intercombination line of such systems has been proposed or carried out for clocks \cite{ido_precision_2005, olson_ramseyborde_2019,hilton_demonstration_2025,fartmann_ramseyborde_2025}, magnetometers \cite{rathod_magnetometry_2015, nanarong_quantum_2025} and thermometers \cite{truong_atomic_2015}.  Thermal atomic beams of alkaline earth atoms are also a promising platform for ultra-narrow-linewidth super-radiant lasers \cite{liu_rugged_2020,jager_superradiant_2021,tang_prospects_2022, fama_continuous_2024}.  

The experimental complexity of generating narrow linewidth (sub-kilohertz) laser light with its frequency referenced to the atomic intercombination transition remains an obstacle to the wider deployment of these systems.  Current methods rely on stabilization to bulk ultrastable cavities, either directly or via a frequency comb \cite{koller_transportable_2017,bothwell_deployment_2025}, or the use of heat pipes or atomic beams based on bulk vacuum hardware for direct atomic spectroscopy \cite{li_narrow-line_2004, jayakumar_dualaxis_2015}.  
The recent development of low noise lasers based on nanophotonic resonators \cite{isichenko_sub-hz_2024, loh_optical_2025} offers a promising path toward compact, manufacturable laser sources for interrogating the intercombination lines without the need for bulk ultrastable cavities.  Pairing these integrated lasers with a compact, manufacturable alkaline earth atomic source, such as a microfabricated vapor cell \cite{pate_microfabricated_2023}, could aid the development of compact quantum sensors based on alkaline earths.  

For these applications, it is desirable to have a narrow spectroscopic feature, free from environmental perturbations that can cause the line to broaden and shift.  In particular, residual gas contamination in vapor cells can result in both shifts and broadening of the spectroscopic line, with typical broadening coefficients on the order of \qty{10}{MHz/Torr} (\qty{7.5}{\mega\hertz\per\hecto\pascal}) \cite{shiga_buffer-gas-induced_2009}. Previous work on alkali metal microfabricated vapor cells indicated that background gas pressure can be reduced to below \qty{100}{mTorr} with the use of non-evaporable getter pumps \cite{hasegawa_effects_2013, newman_highperformance_2021}, enabling the observation of the two-photon transition in Rb at 778 nm with a linewidth of \qty{1.4}{\mega\hertz}.  A background pressure below 1 millitorr would be required to fully resolve the \qty{7.5}{kHz}-wide natural linewidth of the Sr intercombination transition   \cite{drozdowski_radiative_1997}. %

Our previous work \cite{pate_microfabricated_2023} demonstrated sub-Doppler spectroscopy on the strong $^1$S$_0$ to $^1$P$_1$ transition at \qty{461}{nm} in a microfabricated Sr vapor cell.  The transition's broad \qty{30}{\mega \hertz} natural linewidth limited our measurement resolution of the residual broadening due to background gas in the vapor cell to about \qty{3}{MHz}.  Here we report a 100 kHz linewidth for the $^1$S$_0$ to $^3$P$_1$ intercombination line of $^{88}$Sr at 689 nm in a microfabricated Sr vapor cell. This represents a factor of 10 reduction in the observed atomic linewidth compared to the narrowest optical-transition linewidths observed to date in alkali-atom microfabricated cells, and puts an upper bound on the residual gas pressure in the Sr vapor cell of \qty{10}{\milli Torr} (\qty{1.3}{\pascal}).

\section{Experimental Methods}
\begin{figure*}[th]
    \centering
    \includegraphics[width=1.0\linewidth]{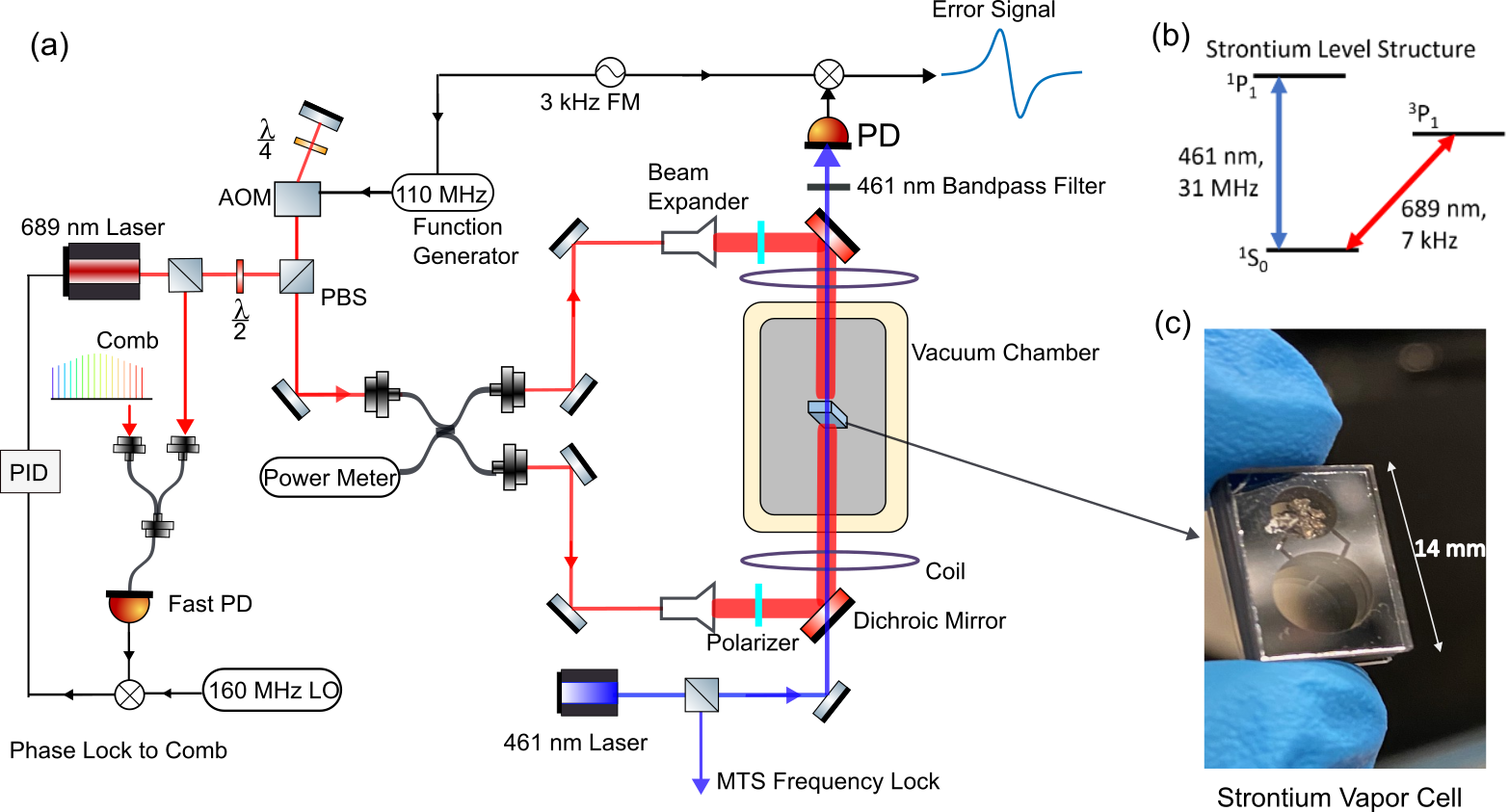}
    \caption{(a) Shelving spectroscopy schematic. MTS: modulation transfer spectroscopy. LO: local oscillator. PD: photodetector (b) Energy levels related to the shelving spectroscopy. (c) Microfabricated Sr vapor cell.   }
    \label{fig:fig1}
\end{figure*}

Figure 1(c) shows a photograph of a microfabricated Sr vapor cell.  The fabrication process is detailed in Ref. \cite{pate_microfabricated_2023} and summarized here. Using deep reactive ion etching, two chambers connected by angled baffles are etched into a Si chip with 9 x 14 x \qty{4.4}{mm^3} external dimensions. A borosilicate glass window with a \qty{20}{nm}-thick layer of Al$_2$O$_3$ deposited on it using atomic layer deposition (ALD) is anodically bonded to one side of the Si frame. A piece of Sr metal is placed into the smaller reservoir chamber in an argon environment. A second window is placed on top of the  resulting chip pre-form and the full vapor cell stack is then quickly transferred into a wafer bonder to avoid oxidizing the Sr. The wafer bonder chamber is pumped down to below \qty{e-5}{\hecto\pascal} and a final anodic bonding step seals the second window onto the chip to complete the fabrication process.

We perform sub-Doppler shelving spectroscopy \cite{manai_shelving_2020}  on the $^1$S$_0$ to $^3$P$_1$ transition at \qty{689}{nm} and measure the linewidth as a function of laser power.  Figure 1(a) shows an overview of the strontium spectroscopy setup, which consists of a vacuum oven to heat the Sr vapor cell, a narrow linewidth spectroscopy laser at \qty{689}{nm} to interrogate the atoms, and a detection laser at \qty{461}{nm} to measure the excitation to the $^3$P$_1$ with high signal to noise ratio (SNR). Figure 1(b) shows the energy levels for the pair of transitions used in the shelving spectroscopy. Since the natural lifetime of the $^3$P$_1$ state, $\tau = 1/\Gamma_{689} = 1/(2\pi \times \qty{7.5}{\kilo \hertz})$ is long compared to other experimental time scales, excitation to the this state can be observed as a reduction in absorption on the strong $^1$S$_0$ to $^1$P$_1$ transition at \qty{461}{nm}. This improves the observed signal strength by roughly $\Gamma_{461} / \Gamma_{689} \approx 4000$, where $\Gamma_{461}$ and $\Gamma_{689}$ are the natural linewidths of the \qty{461}{nm} and \qty{689}{nm} transitions, respectively \cite{manai_shelving_2020}.

\begin{figure*}[htb]
    \centering
    \includegraphics[width=1.0\linewidth]{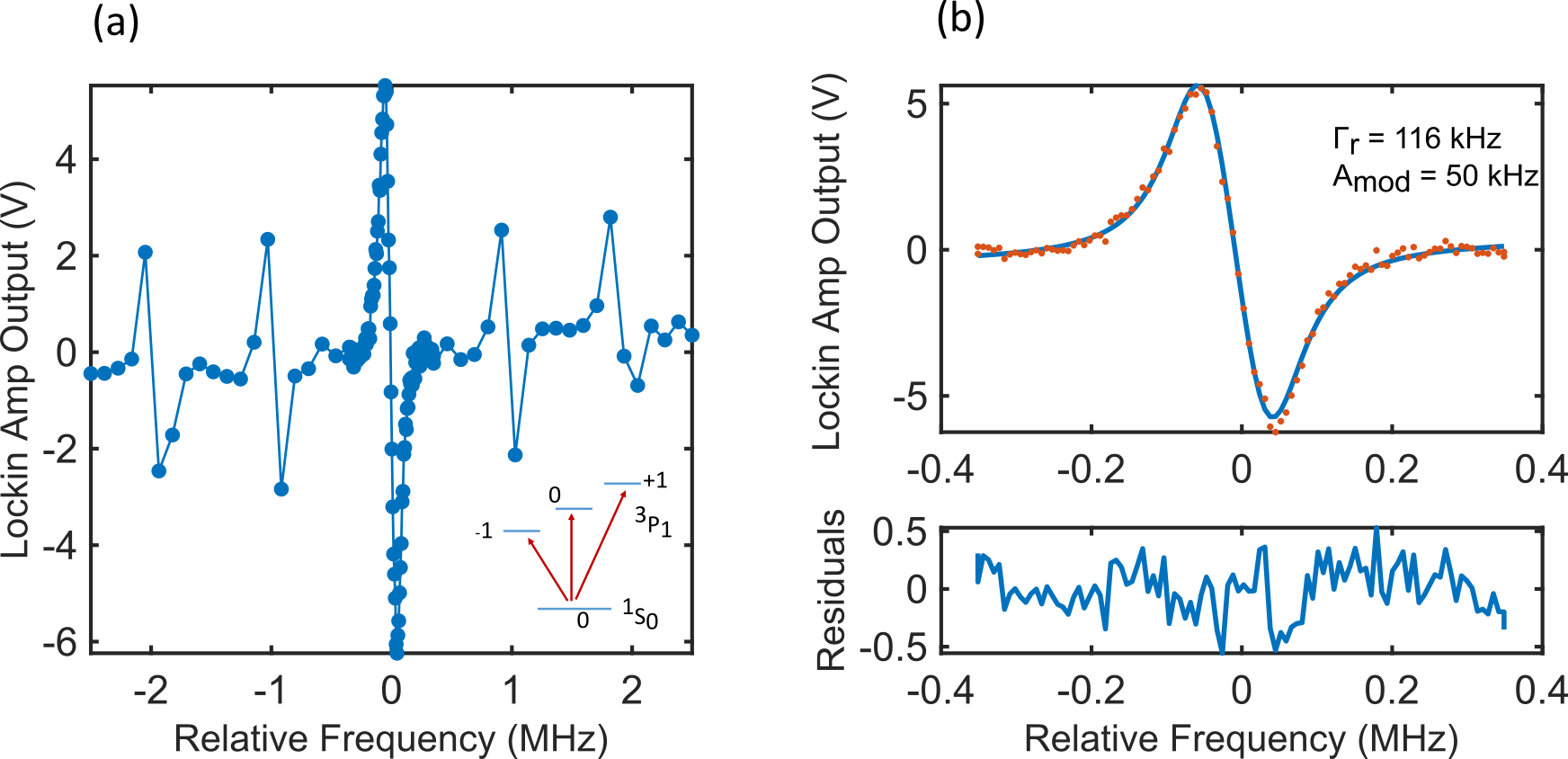}
    \caption{Shelving spectrum for a  beam waist  w = \qty{4.4}{mm} and power of \qty{53}{\micro W} for the 689 nm probe beam. (a) The five dispersive features correspond to 0 to $m_J=$ 1, 0 -1, and their cross-over. Inset shows the related $m_J$ states. (b) Zoom in on the 0 - 0 feature in the center and fit to a model lineshape function. Each point in the spectrum corresponds to an averaging time of \qty{2}{s}. 
    }
    \label{fig:fig2}
\end{figure*}

We typically operate our shelving spectroscopy at a Sr vapor pressure that corresponds to about \qty{50}{\%} absorption of the \qty{461}{nm} detection beam through the \qty{3}{mm} long vapor cell.  This corresponds to a strontium atomic density of $\approx$\qty{e11}{cm^{-3}} and a cell operating temperature of \qty{350}{\degreeCelsius}.  To reach this temperature range, the cell is placed inside a vacuum oven  \cite{pate_microfabricated_2023} at a pressure below \qty{1}{\hecto\pascal} to prevent convective heat loss. Note that the Sr vapor cell was already sealed under pressure much lower than the oven pressure. A pair of field coils placed outside the the vacuum chamber generates a magnetic field along the laser beam propagation axis that, in conjunction with the earth's magnetic field and the field generated by the heating element, defines the atomic quantization axis. During the course of the measurement campaign over one year, the vapor cell was temperature cycled between \qty{25}{\degreeCelsius} and \qty{350}{\degreeCelsius} more than 50 times. At the higher temperature, we achieve a cell operational lifetime of more than \qty{1000}{h}, although there is some degradation in the cell transmission.

The spectroscopy laser system at \qty{689}{nm} consists of an external cavity diode laser (ECDL) stabilized to a self-referenced optical frequency comb. The free running \qty{689}{\nano\meter} laser experiences frequency jitter that broadens its linewidth to about \qty{1}{\mega\hertz}.  To narrow its linewidth for probing the Sr intercombination line we phase lock the spectroscopy laser to a tooth of the frequency comb using a high-bandwidth digital phase-frequency detector and fast PID servo. The repetition rate of the optical frequency comb is stabilized via an optical-phase-lock-loop to a low-noise \qty{1550}{nm} laser, which is in turn  locked to an ultra-stable optical cavity to suppress the short-term fluctuation of the frequency comb.  We monitor the ultra-stable optical cavity drift by measuring the repetition rate of the frequency comb using a maser-referenced phase noise analyzer.  To acquire spectra, the frequency of the spectroscopy laser is stepped across the atomic transition by adjusting the local oscillator frequency of the phase lock between the spectroscopy laser and the frequency comb.

The light is then coupled into a polarization maintaining optical fiber and split into two arms. The light in each arm is roughly power balanced and expanded to a $1/e^2$ waist radius of up to \qty{4.4}{mm}. The two counter-propagating red beams are linearly polarized before entering the vapor cell; their polarization angles are matched by aligning the angles of the two polarizers. The beam pointing misalignment between the two counter-propagating beams is minimized by maximizing the power meter reading of the return port at the input of fiber splitter.  

The blue readout light at \qty{461}{nm} is provided by a second ECDL.  A fraction of the laser power is sampled and used to lock the frequency of the \qty{461}{nm} laser to the peak of the $^1$S$_0$ to $^1$P$_1$ transition using modulation transfer spectroscopy \cite{shirley_modulation_1982} in a second microfabricated Sr vapor cell (not shown in  Fig. 2).  The blue beam is overlapped with the center of the larger red beam using a dichroic mirror. The blue laser power is \qty{20}{\micro W} with a beam waist of \qty{0.25}{mm}. After passing through the vapor cell, the blue detection beam is separated from the red probe beam by a second dichroic mirror and its transmitted power is measured by a photodetector.  The spectroscopy signal is generated using frequency modulation (FM) spectroscopy with a double-pass acousto-optic modulator (AOM) on the red transition, which is also used to stabilize the power of the 689 nm probe beam by feedback to the amplitude of the RF drive signal. The probe beam power can also be varied by setting the amplitude of the RF drive signal. The AOM drive frequency is modulated at a rate of \qty{3.3}{kHz} with an amplitude of $A_{mod}/2$ = \qtyrange{5}{25}{kHz}, where $A_{mod}$ is the center-to-peak FM amplitude seen by the atoms.   

\section{Results and Discussion}

Figure \ref{fig:fig2} shows the shelving spectroscopy signal for the \qty{689}{nm} transition in the presence of a \qty{0.1}{mT} magnetic field over a wide scan range of \qty{5}{MHz}.  The orientation of the spectroscopy beam's linear polarization relative to the applied magnetic field allows for driving all the $\sigma^+$, $\sigma^-$, and $\pi$ transitions.  As a result, we observe five spectral features, corresponding to the magnetically sensitive $\sigma^+$, $\sigma^-$ transitions, the magnetically insensitive $\pi$ transition, and their cross-over features.  The central feature includes contributions from the overlapping of the $\pi$ transition and the $\sigma^+-\sigma^-$ cross-over transition.  

In Fig. \ref{fig:fig2}(b), we fit the central feature from Fig. \ref{fig:fig2}(a) with a model line shape.   The model line shape is the sum of the derivative of a FM-broadened Lorentzian peak \cite{wahlquist_modulation_1961} and a linear slope and offset to account for the slowly changing background from the broad Doppler feature upon which the narrow line sits.   When extracting the residual Lorentzian linewidth, $\Gamma_\textrm{r}$, from the fit,  the modulation parameter is set to the known frequency modulation amplitude, $A_\textrm{mod}$, applied to the \qty{689}{nm} probe beam by the AOM.  The residual linewidth can include contributions from the natural linewidth, $\Gamma_0 = $\qty{7.5}{kHz},  probe laser linewidth, transit time broadening, power broadening, and other inhomogeneous line broadening mechanisms, such as residual Doppler broadening from imperfect laser alignment.

\begin{figure}[tb]
    \centering
    \includegraphics[width=1.0\linewidth]{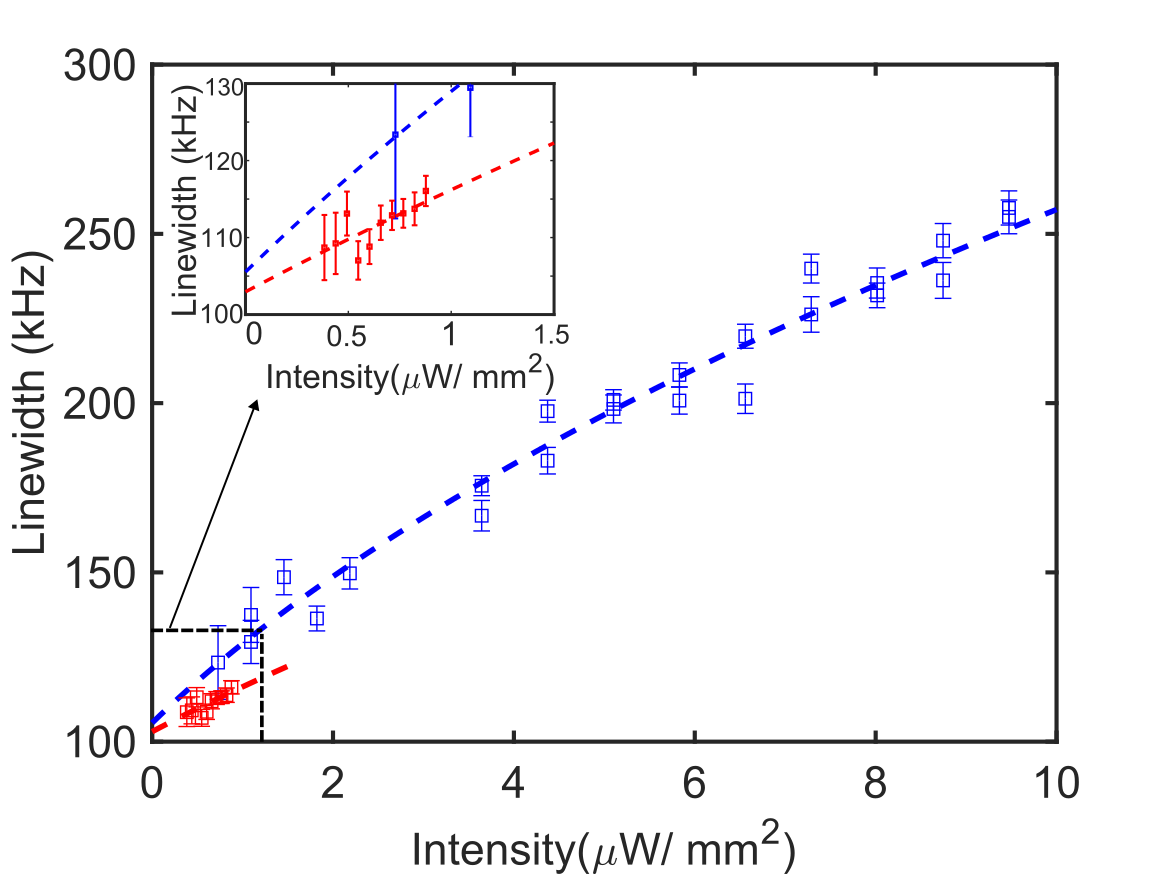}
    \caption{Power broadening of the shelving spectrum.  Error bars represent one $\sigma$ confidence intervals from the fitted lineshape model. Inset zooms in on the linewidths of the bigger beam. Blue squares: beam waist of $w = $ \qty{1.7}{mm}; red squares: $w =$~\qty{4.4}{mm}; dashed lines are fits to data.  }
    \label{fig:fig3}
\end{figure}

Figure \ref{fig:fig3} shows the measured residual Lorentzian linewidth, $\Gamma_r$ as a function of probe beam intensity for two different probe beam $1/e^2$ waist radii of $w = \qty{1.7}{mm}$ and \qty{4.4}{mm}.  We use probe beam intensities ranging from about \qty{0.5}{\micro W.\milli m^{-2}} to \qty{10}{\micro W.\milli m^{-2}}, corresponding to $I \approx$ $15 I_\textrm{sat}$ to $300 I_\textrm{sat}$, where $I_\textrm{sat}=$ \qty{30}{\nano W.\milli m^{-2}}.  The dashed lines represent fits to the data of the form $\Gamma_\textrm{r} = \Gamma' (1 + I/I_0)^{1/2}$, where $\Gamma'$ is the zero-power linewidth, and $I_0$ is a normalization factor.  The fitted zero-power linewidths are \qty{105(1)}{kHz} and  \qty{103(1)}{kHz}, for the $w=\qty{1.7}{mm}$ and \qty{4.4}{mm} beams, respectively. The remaining additional broadening of $\approx$\qty{100}{kHz} beyond the natural linewidth is similar to those observed in tabletop vapor cell experiments\cite{li_narrow-line_2004,shiga_buffer-gas-induced_2009, manai_shelving_2020}. The broadening  contributions from Sr-Sr collisions \cite{crane_measurement_1994}, laser linewidth, and the recoil shift are all expected to be below \qty{5}{kHz}.  Since the atoms detected by our shelving spectroscopy setup are located in the blue beam, at the center of our vapor cell, the typical length an atom travels before being detected is in on the order of the probe beam radius.  We estimate that the transit time broadening and the residual Doppler broadening both contribute to the linewidth at the tens of kilohertz level. 

The disagreement between the experimental data at the two beam sizes in Fig. \ref{fig:fig3} is most likely due to  inaccuracies in the measurement of the two beam waists. Nevertheless, the fitted value of the zero power linewidth does not depend on the size of the beam waist and should not be affected by the corresponding uncertainty in this quantity. Since the fitted zero-power linewidths are consistent despite the beam waist difference, it is likely that the transition linewidth is dominated by broadening mechanisms other than transit time broadening, such as the residual Doppler broadening and broadening due to the laser linewidth mentioned above.

Assuming the entire 100 kHz broadening comes from the pressure broadening due to residual Ar gas, we estimate an upper bound of the residual gas pressure of \qty{10}{\milli Torr} (\qty{1.3}{\pascal})\cite{shiga_buffer-gas-induced_2009} in our microfabricated vapor cell. 

\section{Conclusion}
In conclusion, we demonstrate \qty{100}{kHz}  linewidth of the 689 nm Sr intercombination line in a microfabricated  vapor cell, nearly an order of magnitude smaller than those achieved in its alkali-atom counterparts. The \qty{100}{kHz} broadening puts an upper limit on the residual gas pressure in these vapor cells of \qty{10}{mTorr}. The measured linewidth in our microfabricated vapor cell shows no obvious degradation compared with the linewidths in other tabletop vapor cell experiments \cite{li_narrow-line_2004,shiga_buffer-gas-induced_2009, manai_shelving_2020}. Our results pave the way for a new generation of scalable, compact optical frequency standards, as well as reference cells for laser locking in optical lattice clocks and some quantum computing platforms based on alkaline earth atoms. \\

\noindent\textbf{Funding.} National Institute of Standards and Technology.\\

\noindent\textbf{Acknowledgements.} The authors thank Travis Briles and Zheng Luo for helpful discussions. Dr. Yang Li acknowledges support under the Professional Research Experience Program (PREP), funded by the National Institute of Standards and Technology and administered through the Department of Physiscs, University of Colorado, Boulder.

\bibliography{references.bib}

\end{document}